\documentclass[reprint,amsmath,amssymb,aps]{revtex4-2}
\usepackage{amsmath}
\usepackage{tipa}
\usepackage{bbm}
\usepackage{txfonts}
\usepackage{graphicx}
\usepackage{dcolumn}
\usepackage{bm}
\usepackage{amssymb}
\usepackage{latexsym}
\usepackage{color}
\usepackage{graphicx}
\usepackage{floatrow}
\usepackage[colorlinks=true,linkcolor=blue,citecolor=blue,urlcolor=blue,]{hyperref}
\usepackage[doipre={doi:~}]{uri}

\begin{document}

\bibliographystyle{apsrev4-2.bst}

\title{Manifold formation and crossings of ultracold lattice spinor atoms in the intermediate interaction regime}
\author{Xue-Ting Fang}
\author{Zheng-Qi Dai}
\author{Di Xiang}
\author{Shou-Long Chen}
\author{Shao-Jun Li}
\author{Xiang Gao}
\author{Qian-Ru Zhu}
\author{Xing Deng}
\author{Lushuai Cao}\email[E-mail: ]{lushuai\_cao@hust.edu.cn}
\author{Zhong-Kun Hu}

\affiliation{MOE Key Laboratory of Fundamental Physical Quantities Measurement $\& $
Hubei Key Laboratory of Gravitation and Quantum Physics, PGMF and School of Physics,
Huazhong University of Science and Technology, Wuhan 430074, P. R. China}

\date{\today}

\begin{abstract}
Ultracold spinor atoms in the weak and strong interaction regime have received extensive investigations, 
while the behavior in the intermediate regime is less understood.
We numerically investigate ultracold spinor atomic ensembles of finite size 
in the intermediate interaction regime,
and reveal the evolution of the eigenstates from the strong to the intermediate regime.
In the strong interaction regime, it has been well known that the low-lying eigenenergy spectrum presents
the well-gaped multi-manifold structure, and the energy gaps protect the categorization of the eigenstates.
In the intermediate interaction regime, it is found that the categorization of the eigenstates is preserved, 
and the eigenenergy spectrum become quasi-continuum, with different manifolds becoming overlapped. 
The overlapping induces both direct and avoided crossings between close-lying
manifolds, which is determined by the combined symmetries of the eigenstates involved in the crossing.
A modified t-J model is derived to describe the low-lying eigenstates in the intermediate regime, which can capture the formation
and crossings of the manifolds. State preparation through the avoided crossings is also investigated.

\end{abstract}

\pacs{37.25.+k, 03.75.Dg, 04.80.Cc}

\maketitle

\section{Introduction}
Spinor quantum gases normally refer to the ultracold atoms, of which the internal states are taken as the spin degree of freedom \cite{kawaguchi2012,stamper2013}.
Ultracold spinor gases have become an important platform in various fields, such as the quantum magnetism, quantum phase transition,
and topological excitations. Magnetic phases \cite{stenger1998,ciobanu2000,sadler2006,huh2020} and associated phase transitions \cite{greiner2002,greiner2002,zwierlein2006,roch2008,gemelke2009,murthy2015,williamson2016,williamson2017,fujimoto2018,guo2021} have been investigated on the spinor atomic platform, and various topological excitations such as 
vortices \cite{flayac2010,montgomery2013,ueda2014} and monopoles \cite{ruostekoski2003,pietila2009,ray2015} have also been theoretically
proposed and experimentally realized with spinor atoms. Besides the fundamental interests, spinor atoms also provide a promising platform
for quantum simulations \cite{cazalilla2009,gorshkov2010,messio2012,zhang2014,yang20201,yang20202,dai2017,sun2021} and quantum metrology \cite{luo2017,zou2018,liu2022,pezze2018,gross2010}. The simulations of, e.g., topological systems \cite{yang20202,dai2017,sun2021} and 
and high energy physics \cite{cazalilla2009,gorshkov2010,messio2012,zhang2014} have realized with lattice spinor atoms. The entangled and/or squeezed spinor atoms \cite{kitagawa1993,MA201189,hosten2016} are well recognized as important source in quantum measurements.

The interaction between spinor atoms plays a key role in the above mentioned studies and applications. Theoretical tools have been developed
for the spinor atoms in the weak and strong interaction regimes, which have provided deep insights into the stationary and dynamical behavior
in the corresponding interaction regimes. In the weak interaction regime, the spinor atoms are in the condensate state, and well described by the single-mode approximation (SMA), in which the condensate state is assumed
taking the same spatial wavefunction for all spin states. The ground state \cite{yi2002,koashi2000,yi2004,pu1999,ho1998,xu2019,chang2007,evrard20211,evrard20212,guan2021,leanhardt2003,yi2006}, excitation \cite{pu1999,ho1998,xu2019,PhysRevLett.90.250403} and
the dynamical properties \cite{pu2000,chang2005,chang2007,jie2020} of the spinor condensate have been revealed under SMA, and the validity of SMA is also investigated \cite{pu1999,jie2020}. 
In the Tonks-Girardeau (TG) regime \cite{kinoshita2004,palzer2009,paredes2004,vignolo2013}, the Bose-Fermi mapping is manifested as a good analytical tool
for the spinor atoms. In the strong interaction regime, analytical ansatz based on perturbation treatment with the Bose-Fermi mapping has been developed
and provided good understanding for half-spin \cite{deuretzbacher2014,volosniev2014,levinsen2015,massignan2015,volosniev2015,marchukov2016,yang2015,yang2016,barfknecht2017,deuretzbacher2017,alam2021} and integer-spin \cite{yang20161,jen2017,liu2017}systems, which reveals the multi-manifold structure in the eigenenergy spectrum and deduces effective Heisenberg
model for each manifold. The strongly interaction spinor atoms have become a promising test bed for the strong correlation induced phase
\cite{levinsen2015,massignan2015,volosniev2015,liu2017} transition and dynamical processes \cite{deuretzbacher2014,barfknecht2017} .

\begin{figure}[b]
\includegraphics[trim=30 20 60 30,width=1\textwidth]{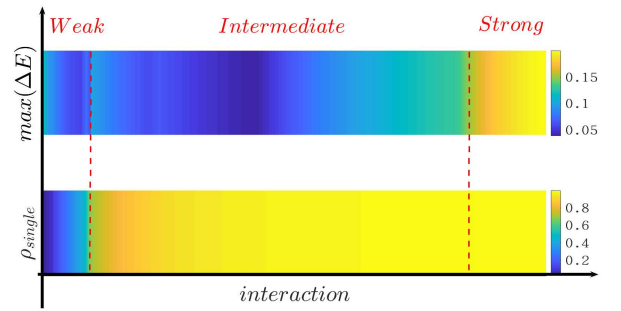}
\caption{\label{fig:1}Sketch of the transition between the weak, intermediate and strong interaction regime,
in terms of the appearance of the energy gap (upper bar) and the total probability of the single-occupation basis states (lower bar).
The appearance of the energy gap is revealed by the maximum energy difference between close-lying eigenenergies and
the results are calculated with the finite $\left( {3 \uparrow 3 \downarrow 1h} \right)$ system.}
\end{figure}

In between the weak and strong interaction regime, there lies a wide intermediate regime, in which the behavior of the spinor atoms is less well
understood. Concerning the lattice spinor atoms, the transition between the three interaction regimes can be indicated by two criterions,
namely the continuity of the eigenenergy spectrum and the local density fluctuation in the lattice. As sketched in Fig. \ref{fig:1},
the weak interaction regime is characterized by the (quasi-)continuum spectrum and large occupation fluctuation illustrated by the
probability of the single-occupation basis, whereas in the strong interaction regime the spectrum becomes gapped and occupation is 
strongly suppressed, as the single-occupation states become dominant. The intermediate regime behaves as hybridization of the weak 
and strong interaction regime,
with the (quasi-)continuum spectrum and the domination of the single-occupied basis in the low-lying eigenstates. This hybridization 
suggests that the intermediate regime could provide novel phenomena different from those in the weak or strong interaction regimes.

In this work, we perform numerical simulations on finite ultracold spinor atoms confined in one-dimensional optical
lattices, and our numerical simulation reveals the transition from the strong to the intermediate regime.
In the strong interaction regime, it has been known that the eigenenergy spectrum presents well-gaped multi-manifold structure,
and the eigenstates can be correspondingly categorized into different manifolds \cite{volosniev2014,yang2015}, which is protected by the energy gaps against
the inter-manifold coupling. In the intermediate regime, for one thing, the energy spectrum becomes quasi-continuum, with the
energy gaps vanished. For another, the eigenstates can still be categorized into different manifolds, even without the protection
of the energy gap between different manifolds. It turns out that the categorizability of the eigenstates in the intermediate
regime is attributed to the competition between the finite energy spacing between close-lying eigenstates and the inter-manifold
coupling, of which the finite energy spacing dominates over the inter-manifold coupling and maintains the categorization of the
eigenstates. The quasi-continuum energy spectrum gives rise to the overlapping between different manifolds, 
which leads to rich energy level crossings between different
manifolds. The avoided crossings can be explored for state preparation and manipulations.
We derived a modified t-J model to describe the lattice spinor atoms in the intermediate regime, and the t-J model well
explains the manifold structure preservation in the intermediate regime, and reveals the influence of the spin and spatial related
symmetries on determining whether the energy level crossing is a direct or avoided one. The dynamical magnetization through
interaction quench between avoided crossing points is also numerically demonstrated.

The manuscript is organized as follows: In Sec.\ref{II} we present the setup under consideration, and present the derivation of our low-energy effective Hamiltonians. In Sec.\ref{III} we show the energy spectrum obtained from the numerical method and energy level crossings between different manifolds. Finally, a brief discussion and conclusion are given in Sec.\ref{IV}.

\section{Set up and effective Hamiltonians}\label{II}
We consider the bosonic spinor atoms confined in the one-dimensional optical lattices. Two internal states of the atoms are chosen to span the spin degree of freedom, and are denoted as $\left|  \uparrow  \right\rangle $ and $\left|  \downarrow  \right\rangle $,
that’s to say we focus on the effective spin-1/2 systems in this work. 
The lattice spinor atomic system is subjected to the Hubbard Hamiltonian, as
\begin{equation}
\begin{aligned}
H =  - t\sum_{ < i,j > ,\sigma } {\hat a_{i,\sigma }^\dag {{\hat a}_{j,\sigma }}}  
+ U\sum_{i,\sigma ,\sigma '} {\hat a_{i,\sigma }^\dag \hat a_{i,\sigma '}^\dag {{\hat a}_{i,\sigma '}}{{\hat a}_{i,\sigma }}} ,
\end{aligned}
\end{equation}
where the ${\hat a_{i,\sigma }}\left( {\hat a_{i,\sigma }^\dag } \right)$ indicates the annihilation (creation) operators of an atom of spin state $\sigma$ in the $i$-th site. $\left\langle {i,j} \right\rangle $ denotes the summation over the nearest neighbors in the lattice. The first and second terms of $H$ refer to the nearest neighbor hopping and
the on-site interaction of the atoms, respectively,
of which $t$ and $U$ are the spin-independent tunneling strength and contact interaction strength. 

In the numerical simulations, finite spin-balanced atomic ensembles with $N$ spin-up and $N$ spin-down atoms confined
in a lattice with $2N+1$ sites are considered, i.e. the cases of single-hole filling. 
The numerical simulations are performed in the complete Hilbert space spanned by all the basis states,
including both the single- and multiple-occupation states, through the method of multi-layer multi-configuration
time-dependent Hartree method for mixtures of arbitrary species \cite{cao2013,kronke2013,cao2017}. The analysis of the numerical results, however, 
is carried out within the truncated Hilbert space spanned by the single-occupation states, 
since we mainly focus on the low-lying eigenstates, which are dominated by the single-occupation states. 
The validity of the truncation of the low-lying eigenstates to the single-occupation basis is verified by
checking the probability of the single-occupation states in the eigenstates of interests.

\begin{figure*}[t]
\includegraphics[trim=50 20 30 30,width=0.8\textwidth]{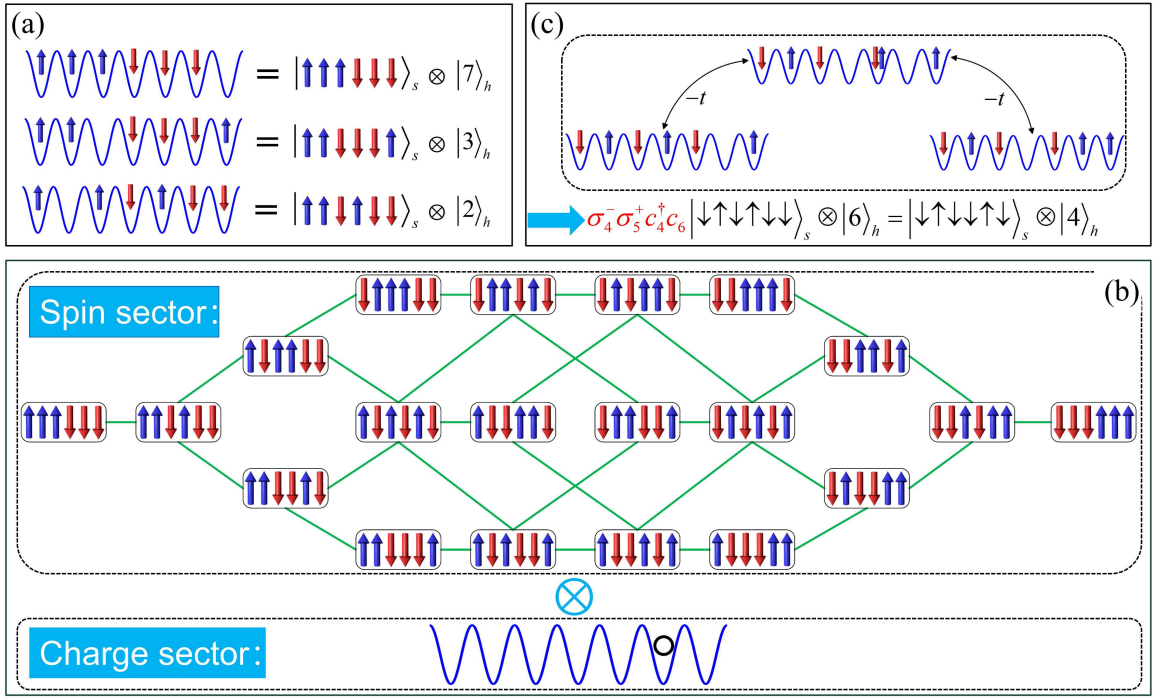}
\caption{\label{fig:2}Illustration of spin-charge separation in the $\left( {3 \uparrow 3 \downarrow 1h} \right)$ system:
(a) basis transformation from the original to the spin-charge basis, (b) the configuration space of the spin and charge sector,
and (c) the high-order tunneling process of the simultaneous spin flipping and next-nearest-neighbor hopping of the hole via
the intermediate double occupation states.}
\end{figure*}
In the analysis, the truncated Hilbert space spanned by the single-occupation basis is transformed to the direct product of
the charge sector and spin sector. The charge sector is composed of the configuration space of
a single hole hopping in the $2N+1$-site lattice and the spin sector corresponds to  a spin chain of $2N$ spins.
The single-occupation basis state 
$\hat{a}^\dag_{i_1,\sigma_1}\hat{a}^\dag_{i_2,\sigma_2}\cdots\hat{a}^\dag_{i_{2N},\sigma_{2N}}|Vac\rangle$, 
with $i_1<i_2<\cdots<i_{2N}$, can be transferred to the product form of
${\left| {{\sigma _1}{\sigma _2} \cdots {\sigma _{{\rm{2}}N}}} \right\rangle _S} \otimes {\left| i \right\rangle _H}$,
in which ${\left| i \right\rangle _H}$ indicates the location of the hole in the lattice
and ${\left| {{\sigma _1}{\sigma _2} \cdots {\sigma _{2N}}} \right\rangle _S}$ is the Fock configuration of
the spinor atoms in the squeezed space, with the site occupied by the hole removed from the lattice \cite{vijayan2020,hilker2017}.
Figure \ref{fig:2}(a) illustrates the transformation of the single-occupation states to the direct product form, 
in the finite system of three spin-up atoms and three spin-down atoms confined in a lattice of seven sites, 
denoted as $\left( {3 \uparrow 3 \downarrow 1h} \right)$ in the following. 
Figure \ref{fig:2}(b) sketches the configuration space constituted of the spin and charge sectors of the 
$\left( {3 \uparrow 3 \downarrow 1h} \right)$ system.

Following the decomposition of the truncated Hilbert space to the spin and charge sectors, the Hubbard Hamiltonian is transformed to the modified t-J model \cite{emery1990,ogata1991,grusdt2019,grusdt2020z}. 
The modified t-J model of single-hole filling systems reads:

\begin{subequations}
\begin{align}
{H_{\rm{t - J}}} &= {H_{\rm{hole}}} + {H_{\rm{spin}}},\\
{H_{\rm{hole}}} &=  - t\sum_{i = 1}^{M - 1} {\left( {\hat c_i^\dag {{\hat c}_{i + 1}} + h.c.} \right)} ,\\
{H_{\rm{spin}}} &=  - J\sum_{i = 1}^{N - 1} {\left( {{{\vec \sigma }_i} \cdot {{\vec \sigma }_{i + 1}} + 3} \right)\left( {1 - \hat c_{i + 1}^\dag {{\hat c}_{i + 1}} + \frac{{\hat c_i^\dag {{\hat c}_{i + 2}} + h.c.}}{2}} \right)} ,
\end{align}
\end{subequations}
$c_i^{\left( \dag  \right)}$ refers to the annihilation (creation) operator of the hole in the $i$-th site of the lattice, and ${\vec \sigma _j} \equiv \left( {\sigma _j^x,\sigma _j^y,\sigma _j^z} \right)$ is the Pauli matrices of the $j$-th spin in the squeezed space. 
${H_{\rm{hole}}}$ describes the hopping of the hole in the charge sector,
and ${H_{\rm{spin}}}$ in the spin sector indicates the spinor atoms in the squeezed space organized to the Heisenberg spin chain. 
It is worth noticing that in ${H_{\rm{spin}}}$ the spin-spin interaction strength in the spin sector is dependent
on the local occupation and next-nearest neighbor (NNN) correlations of the hole in the charge sector, 
which gives rise to the coupling between the charge and spin sectors. 
The coupling between the two sectors is attributed to the second-order tunneling process, with $J =t^2/U$. 
Figure \ref{fig:2}(c) sketches the action of simultaneous spin flipping and next-nearest-neighbor hopping of the hole in ${H_{\rm{spin}}}$,
which is mediated by double occupation states.
 
Under the condition of $J \ll t$, ${H_{\rm{t-J}}}$ can be decomposed into the leading term ${H_{\rm{manfd}}}$ and the perturbation ${H_{\rm{scatt}}}$, as:
\begin{subequations}
\begin{align}
{H_{\rm{t - J}}} &= {H_{\rm{manfd}}} + {H_{\rm{scatt}}} ,\label{HtJ} \\
{H_{\rm{manfd}}} &= \sum_{\alpha  = 1}^M {{{\left| {{w_\alpha }} \right\rangle }_H}\left\langle {{w_\alpha }} \right|} \Bigg[ {{\varepsilon _\alpha } - J\sum_{i = 1}^{N - 1} {{{\vec \sigma }_i}{{\vec \sigma }_{i + 1}}} } \nonumber \\
& { \times \left( {1 - {{\left| {{w_\alpha }\left( i \right)} \right|}^2} + w_{\alpha}^*\left( i \right){w_\alpha }\left( {i + 2} \right)} \right)} \Bigg],\label{Hmanfd}  \\
{H_{{\rm{scatt}}}} &= \frac{J}{2}\sum_{{\alpha _1} \ne {\alpha _2}}^{} {{{\left| {{w_{{\alpha _1}}}} \right\rangle }_H}\left\langle {{w_{{\alpha _2}}}} \right|} \Bigg[ {\sum_{i = 1}^{N - 1} {{{\vec \sigma }_i}{{\vec \sigma }_{i + 1}}} }  \times \left( {2w_{{\alpha _1}}^*\left( i \right){w_{{\alpha _2}}}\left( i \right)} \right. \nonumber \\
& {\left. { - w_{{\alpha _1}}^*\left( i \right){w_{{\alpha _2}}}\left( {i + 2} \right) - w_{{\alpha _2}}^*\left( i \right){w_{{\alpha _1}}}\left( {i + 2} \right)} \right)} \Bigg],\label{Hscatt}
\end{align}
\end{subequations}
In the above equations, $|w_{\alpha}\rangle_H$ refers to the $\alpha$-th eigenstate of $H_{\rm{hole}}$,
with $w_\alpha(i)$ and $\epsilon_\alpha$ denoting the corresponding eigen-wavefunction and eigenenergy, respectively.
In ${H_{\rm{manfd}}}$, the hole in the charge sector remains to the eigenstates of $H_{\rm{hole}}$, 
and the spin sector turns to the Heisenberg spin chain, with the spin-spin interaction dependent on the wavefunction of 
${\left| {{w_\alpha }} \right\rangle _H}$, which is consistent with the derivation of the site-dependent spin-spin interaction in \cite{volosniev2014,yang2015,yang2016}. 
${H_{\rm{scatt}}}$ refers to the scattering between different $|w_{\alpha}\rangle_H$ in the charge sector.

In the strong interaction regime, the energy difference between $|w_{\alpha}\rangle_H$ is much stronger than the scattering strength of ${H_{\rm{scatt}}}$,
which prevents the coupling between different $|w_{\alpha}\rangle_H$. The eigenstates of ${H_{\rm{manfd}}}$, that's 
$\left| {\alpha ,\Sigma }\right\rangle \equiv {\left| {{w_\alpha }} \right\rangle _H} \otimes {\left| {\Sigma \left( \alpha  \right)} \right\rangle _S}$,
represent a good approximation of those of ${H_{\rm{t-J}}}$, in which 
${\left| {\Sigma \left( \alpha  \right)} \right\rangle _S}$ refers to the eigenstates of the Heisenberg chain in the spin sector.
$\left| {\alpha ,\Sigma }\right\rangle$ can be categorized into different manifolds with respect to ${\left| {{w_\alpha }} \right\rangle _H}$,
and in the eigenenergy spectrum of ${H_{\rm{t-J}}}$, eigenstates of the same manifold are well localized around the corresponding $\epsilon_\alpha$,
which leads to the well-known multi-manifold structure in the low-lying eigenenergy spectrum in the strong interaction regime of ${H_{\rm{t-J}}}$.
It can be summarized that the categorization of the eigenstates of ${H_{\rm{t-J}}}$ with respect to ${\left| {{w_\alpha }} \right\rangle _H}$
in the strong interaction regime is protected by the energy gaps between different $|w_{\alpha}\rangle_H$, which prevents the coupling between
different ${\left| {{w_\alpha }} \right\rangle _H}$.

In the intermediate interaction regime, however, the coupling between different manifolds
due to ${H_{\rm{scatt}}}$ becomes stronger and cannot be neglected, which affect the categorization of the low-lying eigenstates.
Before proceeding to the details in the intermediate regime,
it is worth paying an attention to the symmetries in ${H_{\rm{manfd}}}$ and ${H_{\rm{scatt}}}$,
which play an important role in the coupling between different manifolds.
In ${H_{\rm{manfd}}}$ , the charge sector is subjected to the space reflection symmetry $\hat T_{rc}$, 
and the spin sector is subjected to both the space reflection symmetry $\hat T_{rs}$ as well as the 
spin flipping symmetry $\hat T_{fs}$. Each $\left| {\alpha ,\Sigma } \right\rangle $ is associated with three
parities $\left( {{T_{rc}},{T_{rs}},{T_{fs}}} \right)$, which correspond to the parity of the space reflection 
symmetry in the charge and spin sector, as well as that of the spin flipping symmetry, 
with ${T_{rc}},{T_{rs}},{T_{fs}} \in \left\{  \pm  \right\}$. The Hamiltonian ${H_{\rm{scatt}}}$, however,  
is invariant under the action of $\hat T_{fs}$ and ${\hat T_{rc}}{\hat T_{rs}}$, but not
the individual action of $\hat T_{rc}$ or  $\hat T_{rs}$. The different symmetry of ${H_{\rm{manfd}}}$ 
and ${H_{\rm{scatt}}}$ leads to that ${H_{\rm{scatt}}}$ can only couple $\left| {\alpha ,\Sigma } \right\rangle $
with the same parity (product) of ${T_{fs}}$ and ${T_{rc}}\cdot{T_{rs}}$.

\section{Numerical results on the manifold formation and crossings}\label{III}
\begin{figure}[t]
\includegraphics[trim=25 5 50 5,width=1\textwidth]{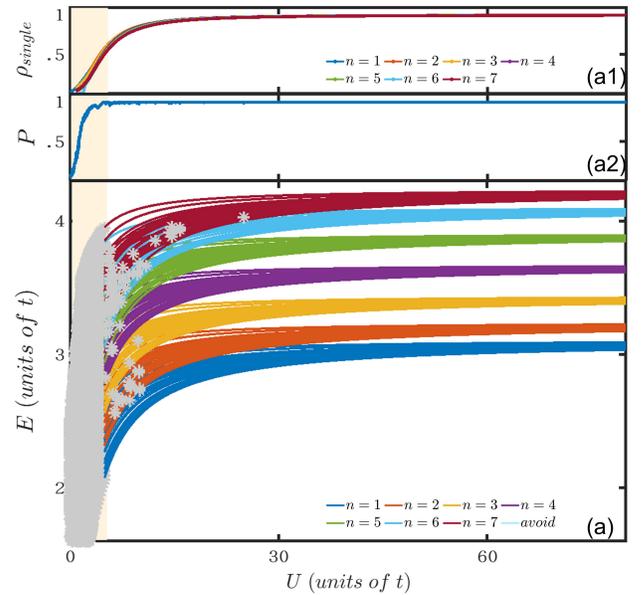}
\caption{\label{fig:3}(a) The eigenenergy spectrum for $\left( {3 \uparrow 3 \downarrow 1h} \right)$. Eigenstates of the same manifold are marked with the same colour, and the eigenenergy of uncategorized eigenstates is shown with gray stars. 
(a1) The probability of the single-occupation basis of the eigenstates in each manifold, and the lowest probability
of all the eigenstates in each manifold is taken for plotting. 
(a2) The ratio of categorizable eigenstates to the total low-lying eigenstates in different interaction regime.}
\end{figure}

We present the numerical results on the eigenenergy spectrum of the $\left( {3 \uparrow 3 \downarrow 1h} \right)$ 
system in Fig. \ref{fig:3}(a). The eigenenergy spectrum are calculated with the interaction strength U scanned over a wide 
interval from the weak to approaching the TG regime, where the eigenenergy spectrum saturates to the fermionization
limit. In the strong interaction regime with the well-gaped multi-manifold structure,
the eigenstates can be categorized into different manifolds, 
and eigenstates in the same manifold are well approximated by $\left| {\alpha ,\Sigma } \right\rangle $
of the same $\alpha$. The multi-manifold structure is protected by the energy gaps, which prevent
the inter-manifold coupling induced by ${H_{\rm{scatt}}}$.
As the interaction decreases from the strong interaction regime, the gaps vanishes and the spectrum becomes
quasicontinuum, which marks the transition from the strong to the intermediate regime.
On the other side of the spectrum, the transition between the intermediate and weak interaction regime is captured
by the decreasing of total probability of the single-occupation basis as U decreases, as shown in Fig. \ref{fig:3}(a1).

In the intermediate interaction regime, where the energy gaps vanish between close-lying manifolds in the spectrum, 
an immediate question is whether the eigenstates can still be categorized into different manifolds with 
no protection of the energy gaps against the inter-manifold coupling.
To directly address this question, we apply the wavefunction categorization to eigenstates in the intermediate regime
\cite{PhysRevLett.126.240402,PhysRevB.100.075102,PhysRevLett.120.257204,Flux.jl-2018,innes:2018},
which is based on the supervised machine learning.
In the supervised categorization, the training set is
chosen from the eigenstates in the strong interaction regime, and the trained network is used to categorize the eigenstates in the
whole interaction interval in Fig. \ref{fig:3}(a). The categorization results are illustrated by the colours in Fig. \ref{fig:3}(a):
The eigenenergies are marked with the same colour, of which the corresponding eigenstates are recognized as belonging to the same
manifold. The eigenenergies marked with grey stars
refer to that the supervised categorization is failed for the corresponding eigenstates with 
the predicted probability below 0.95.
It is found that in the weak interaction regime, all eigenenergies are marked in grey, indicating that the eigenstates in
the this regime cannot be categorized into different manifolds.
In the intermediate regime, however, most eigenenergies are well coloured, and this indicates that in the intermediate regime 
most eigenstates can still be categorized into different manifolds, even without the explicit gaps between close-lying manifolds.
It is also noticeable that a few exceptional uncategorizable eigenstates arise in the intermediate regime, 
which are marked by the sparse grey stars immersed in the well-colored spectrum in this regime.
Figure \ref{fig:3}(a2) quantifies the categorizability in terms of the ratio of categorizable eigenstates to
the total low-lying eigenstates in different interaction regime. 
In the intermediate regime, the ratio is as high as approaching unity, indicating that most
eigenstates can be categorized into different manifolds in this regime. In the weak interaction regime, the ratio presents
a relatively sharp decrease to zero, and the categorization is completely failed . 
Figure \ref{fig:3}(a2) suggests that the categorizability
is manifested as an alternative criterion of the transition between 
the weak and intermediate interaction regime.

\begin{figure}[t]
\includegraphics[trim=25 5 10 10,width=1\textwidth]{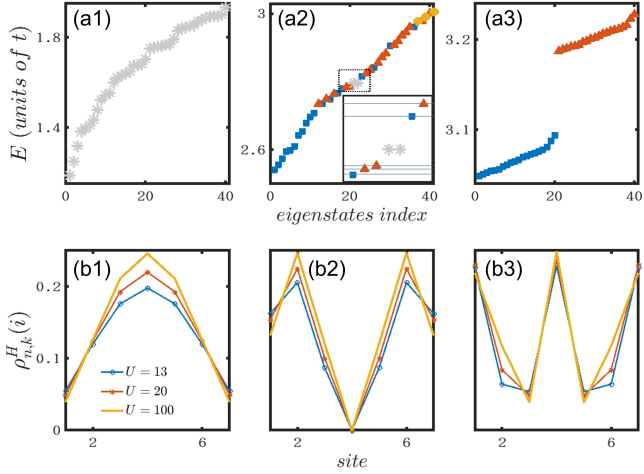}
\caption{\label{fig:4} Eigenenergy spectrum for (a1) $U = 1.6$, (a2) $U = 10$, and (a3) $U = 65$, of which the gray stars refer to
the uncategorizable eigenstates, and the blue squares, red triangles and yellow dots indicate the eigenstates recognized to belonging
to the first, second and third manifold, respectively.
The inset of (a2) zooms in the area marked by dashed box in the main figure, in which the bars located at each categorizable eigenenergy
point measure the inter-manifold coupling. 
(b1), (b2) and (b3) plot the spatial distribution of the hole for eigenstates in the first, second and third manifold, respectively, 
at $U=13$, $U=20$ and $U=100$.}
\end{figure}

Figures \ref{fig:4}(a1-a3) provide a close look at the eigenenergy spectra in the weak, intermediate and strong
interaction regime, respectively,
in which the first forty eigenenergies are shown with the colour indicating the categorization as in Fig. \ref{fig:3}(a).
In the weak interaction regime, as shown in Fig. \ref{fig:4}(a1), the spectrum is quasi-continuum 
and the eigenenergies are marked in grey, indicating that the corresponding eigenstates cannot be categorized. 
Oppositely in the strong interaction regime, the spectrum plotted in Fig. \ref{fig:4}(a3) is well gaped, 
which protects the categorization and the multi-manifold structure.
In Fig. \ref{fig:4}(a2), the spectrum in the intermediate regime becomes quasi-continuum, whereas
most eigenstates are still categorizable and grouped into different manifolds.
The vanishing of the energy gap leads to the overlap of different manifolds in the spectrum.

Figures. \ref{fig:4}(b1-b3) plot the spatial distribution of the hole for eigenstates in the first three manifolds
at different interaction strength in the strong and intermediate regime. The spatial distribution is defined as 
$\rho _{n,k}^H\left( i \right) = \left\langle {n,k} \right|\hat c_i^\dag {\hat c_i}\left| {n,k} \right\rangle$,
with $\left| {n,k} \right\rangle$ indicating the $k$-th eigenstates in the $n$-th manifolds.
It is shown that $\rho _{n,k}^H$ of a given manifold remains qualitatively the same, as the interaction changes from
the strong to the intermediate regime.
Moreover, $\rho _{n,k}^H$ of $n=1,2,3$ resembles the density distribution of 
${\left| {{w_\alpha }} \right\rangle _H}$ with $\alpha=1,2,3$, respectively, which indicates the projection of 
 the corresponding eigenstates in the charge sector dominated by ${\left| {{w_\alpha }} \right\rangle _H}$. 
This demonstrates that each categorizable eigenstate
can be approximated by  $\left| {\alpha ,\Sigma } \right\rangle $ in both the strong and intermediate regime .

In order to illustrate how the categorization is maintained in the intermediate regime,
we zoom in the area around a pair of grey eigenenergies in the main figure of Fig. \ref{fig:4}(a2),
and compare the categorizable and uncategorizable eigenenergies in the insert figure.
It can be seen that the categorizable eigenstates are separated by a non-vanishing energy spacing induced by the finite
size effect, while the uncategorizable pair are almost degenerate. For the categorizable eigenstates,
which are well approximated by $\left| {\alpha ,\Sigma } \right\rangle $, the
${H_{\rm{scatt}}}$-induced inter-manifold coupling is calculated and illustrated by 
the width of the shadowed bar located at the associated eigenenergies.
It can be seen that the energy spacing between the categorizable eigenenergies is much wider than the
width of the shadowed bars, which demonstrates that the non-vanishing energy spacing prevents the inter-manifold coupling 
and maintains the formation of the manifold structure.

\begin{figure}[t]
\includegraphics[trim=35 10 35 20,width=1.1\textwidth]{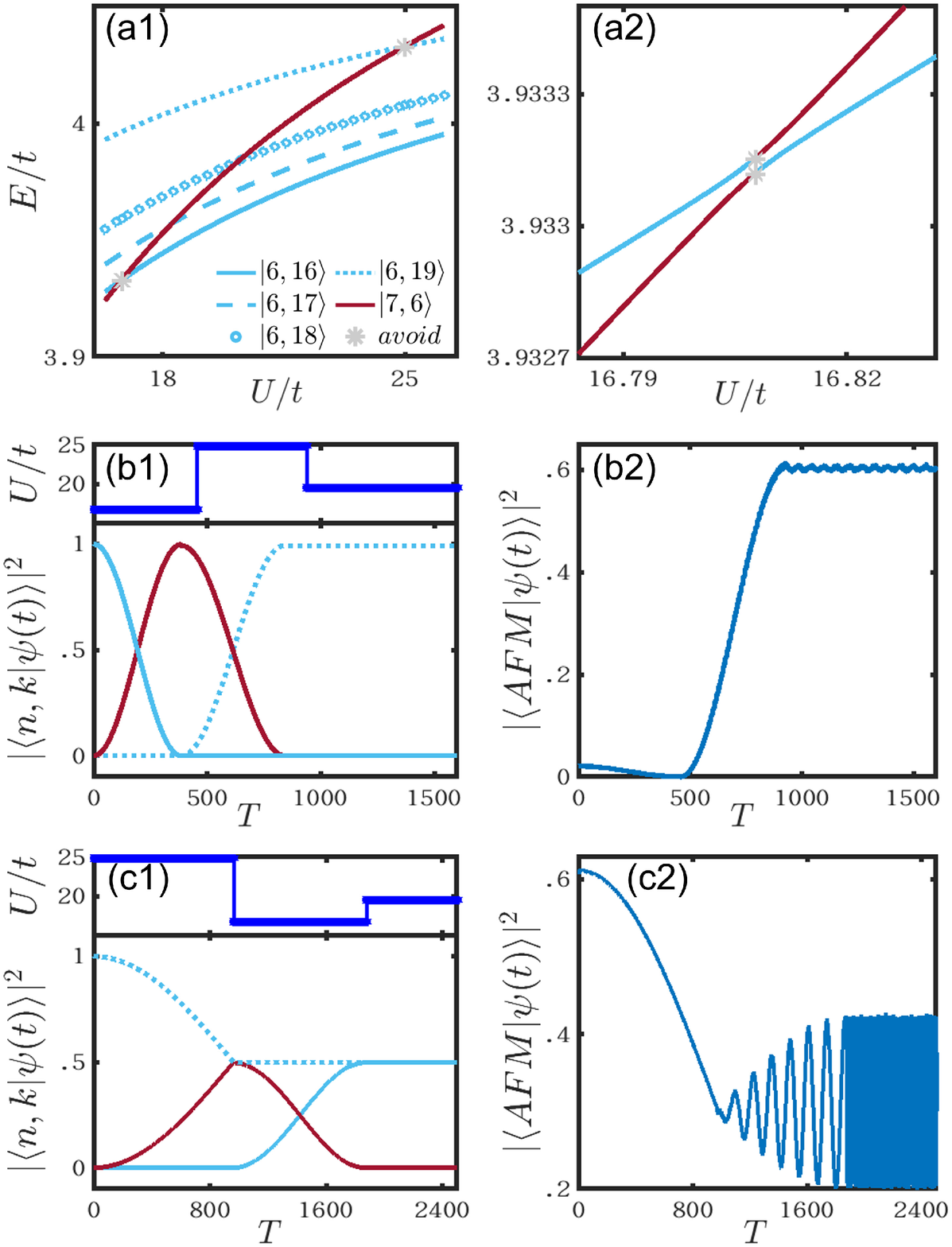}
\caption{\label{fig:5}(a1) Eigenenergy as a function of the interaction strength for $\left| {n  = 6,k  = 16,17,18,19} \right\rangle$
and $\left| {n  = 7,k  = 6} \right\rangle$, around their crossings. The grey stars marks the uncategorized eigenenergies.
(a2) Zoom in around the crossing between  $\left| {n  = 6,k  = 16} \right\rangle$ and  $\left| {n  = 7,k  = 6} \right\rangle$.
(b1-b2) Eigenstate transfer from $\left| {n  = 6,k  = 19} \right\rangle$ to $\left| {n  = 6,k  = 16} \right\rangle$ through two-step
interaction quenches, with (b1) the probability of corresponding eigenstates, and (b2) the probability of the AFM state. Upper panel
of (b1) shows the interaction quench sequence.
(c1-c2) Eigenstate transfer from $\left| {n  = 6,k  = 19} \right\rangle$ to the superposition of 
$\left| {n  = 6,k  = 16} \right\rangle$ and $\left| {n  = 6,k  = 19} \right\rangle$ through two-step
interaction quenches, with (c1) the probability of corresponding eigenstates, and (c2) the probability of the AFM state. Upper panel
of (c1) shows the interaction quench sequence.}
\end{figure}

The major difference between the intermediate and strong interaction regime lies in the uncategorizable eigenstates,
which arises from the accidental degeneracy between $\left| {\alpha ,\Sigma } \right\rangle $ with different $\alpha$.
The accidental degeneracy is also manifested as the crossing between different energy levels in the spectrum.
Figure \ref{fig:5}(a1) takes the crossings between $\left| {n  = 7,k  = 6} \right\rangle$ 
and $\left| {n  = 6,k  = 16,17,18,19} \right\rangle$ for example, where two types of crossings arise: 
One type of crossing is associated with the appearance of uncategorized eigenstates, indicated by the crossings between 
$\left| {n  = 7,k  = 6} \right\rangle$ and $\left| {n  = 6,k  = 16,19} \right\rangle$, and the other is not,
as exemplified by the crossings between $\left| {n  = 7,k  = 6} \right\rangle$ and $\left| {n  = 6,k  = 17,18} \right\rangle$.
Figure \ref{fig:5}(a2) zooms in the crossing between $\left| {n  = 7,k  = 6} \right\rangle$ and 
$\left| {n  = 6,k  = 16} \right\rangle$, and illustrates that the crossing point associated with the uncategorized eigenstates
is manifested as the avoided crossing. The crossings with no uncategorized eigenstates are then of direct crossing.
The coexistence of the two types of crossings is attributed to the symmetry constraints on ${H_{\rm{scatt}}}$ 
as discussed in the previous section, of which ${H_{\rm{scatt}}}$ can only couple $\left| {n,k} \right\rangle$ 
with the same parities of ${\hat T_{fs}}$ and ${\hat T_{rc}}\cdot{\hat T_{rs}}$. 
In Fig. \ref{fig:5}(a2), $\left| {n  = 6,k  = 17} \right\rangle$ and $\left| {n  = 6,k  = 18} \right\rangle$
breaks the constraint on ${\hat T_{fs}}$ and ${\hat T_{rc}}\cdot{\hat T_{rs}}$ with $\left| {n  = 7,k  = 6} \right\rangle$, respectively,
and direct crossing between these eigenstates is observed.

The avoided crossing between a pair of $\left| {n,k} \right\rangle$ can induce the Rabi-like oscillation between the corresponding eigenstates,
and can be explored for state preparations through interaction quenches. Initializing the system in $\left|{n= 6,k= 16}\right\rangle$,
we demonstrate in Figs. \ref{fig:5}(b,c) that the system can be transferred to $\left|{n= 6,k= 19}\right\rangle$ and
the superposition of $\left|{n= 6,k= 16}\right\rangle$ and $\left|{n= 6,k= 19}\right\rangle$,
respectively. The state transfer is accomplished through a two-step interaction quench, mediated by $\left| {n  = 7,k  = 6} \right\rangle$.
Fig. \ref{fig:5}(b1) shows that the system is initially prepare in $\left|{n= 6,k= 16}\right\rangle$, 
and when the interaction is quenched to the bottom avoided crossing
point in Fig. \ref{fig:5}(a1), the system evolves to $\left| {n  = 7,k  = 6} \right\rangle$. 
Upon the system is completely transferred to $\left| {n  = 7,k  = 6} \right\rangle$,
the interaction is quenched to the upper avoided crossing, which finally transfers the system to $\left|{n= 6,k= 19}\right\rangle$.
At each avoided crossing points, the system undergoes a half period Rabi-like oscillation between the corresponding eigenstates.
Given that the projection of $\left|{n= 6,k= 19}\right\rangle$ in the spin sector is dominated by the anti-ferromagnetic state 
$\left|{AFM}\right\rangle=\left(\left|{\uparrow\downarrow\uparrow\downarrow\uparrow\downarrow}\right\rangle+
\left|{\downarrow\uparrow\downarrow\uparrow\downarrow\uparrow}\right\rangle\right)/\sqrt{2}$, this two-step quench 
can be used for antiferromagnetic state preparation.
Similarly, in Fig. \ref{fig:5}(c1), the system can be prepared into the superposition of 
$\left|{n= 6,k= 16}\right\rangle$ and $\left|{n= 6,k= 19}\right\rangle$ through a two-step interaction quench, 
of which the first quench induces a quarter period oscillation between $\left| {n  = 7,k  = 6} \right\rangle$
and $\left| {n  = 6,k  = 19} \right\rangle$, and transfers the system to the superposition of the two eigenstates.
The second quench further transfers the system to the superposition of  $\left| {n  = 6,k  = 16} \right\rangle$
and $\left| {n  = 6,k  = 19} \right\rangle$ by a half period oscillation between $\left| {n  = 6,k  = 16} \right\rangle$
and $\left| {n  = 7,k  = 6} \right\rangle$.
This series of quenches leads to a quantum beating between the antiferromagnetic state and the bi-spinon state
$\left|{BP}\right\rangle=\left(\left|{\uparrow\downarrow\downarrow\uparrow\uparrow\downarrow}\right\rangle+
\left|{\downarrow\uparrow\uparrow\downarrow\downarrow\uparrow}\right\rangle\right)/\sqrt{2}$, which dominates in
$\left|{n= 6,k= 16}\right\rangle$.

\begin{figure}[t]
\includegraphics[trim=25 5 10 5,width=1.07\textwidth]{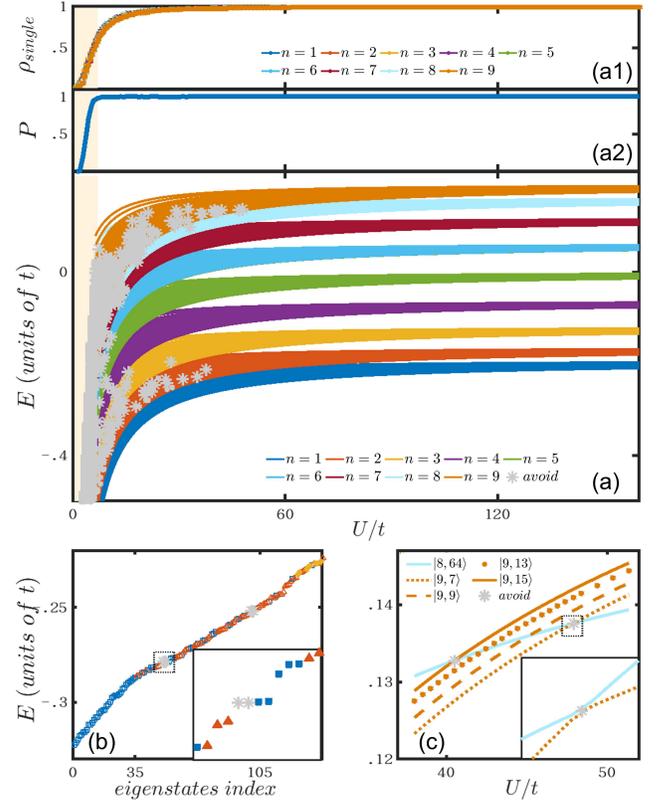}
\caption{\label{fig:6} (a) The eigenenergy spectrum for $\left( {4 \uparrow 4 \downarrow 1h} \right)$. 
Eigenstates of the same manifold are marked with the same colour, and the uncategorized eigenstates are shown with gray stars.
(a1) The lowest occupancy probability of the single-occupation basis of the eigenstates in each manifold. (a2) The ratio of categorizable eigenstates to the total low-lying eigenstates in different interaction regime.
(b) Eigenenergy spectrum at $U=21$, of which gray stars indicate the uncategorized eigenstates, and the blue squares, red triangles
and yellow dots are for the eigenstates of the first, second and third manifold, respectively. 
The inset figure zoom in the dashed box regime in the spectrum, around a pair of uncategorized eigenenergies.
(c) Eigenenergies as a function of interaction strength, for  $\left| {n  = 9,k  = 7,9,13,15} \right\rangle$
and $\left| {n  = 8,k  = 64} \right\rangle$, with the insert zooming in the dashed box marked in the main figure.}
\end{figure}

The formation and avoided crossings of the manifolds in the intermediate regime are
common properties of finite spinor lattices, and can be generalized to larger systems. 
Figure \ref{fig:6} presents the related results in the system of $\left( {4 \uparrow 4 \downarrow 1h} \right)$. 
In Fig. \ref{fig:6}(a), the spectrum as a function of the interaction strength $U$ is shown,
where the well gaped spectrum evolves to a quasi-continuum from the strong to the intermediate interaction regime.
The supervised eigenstate categorization in the whole interaction interval indicates that
the multi-manifold structure is maintained in the intermediate regime, 
which is indicated by the colours of the eigenenergies in the spectrum. 
Figure \ref{fig:6}(a2) quantifies the ratio of the categorizable eigenstates to the total low-lying eigenstates dominated by the single-occupation
states,
and confirms that most eigenstates in the intermediate regime are categorizable.
The overlapping and crossings of different manifolds are then plotted in Fig. \ref{fig:6}(b) and (c), respectively.
The figures confirms that the formation of the manifold structure is attributed to the non-vanishing energy spacing 
due to the finite size effect, and both the direct and avoided crossings show up, of which the avoided crossing
is associated with the uncategorizable eigenstates.

\section{Discussion and Conclusion}\label{IV}
In this work, we numerically investigate the ultracold spinor atoms confined in one-dimensional lattice,
with a focus on the finite lattice systems in the intermediate interaction regime, which are of high relevance
to current experiments. Our investigation reveals the transition from the strong to the intermediate regime, in which
the eigenenergy spectrum becomes quasi-continuum while the eigenstates retain good categorizability into 
different manifold. The formation of the manifold structure in the intermediate regime can be attributed to the
finite size effect, which is normally taken as marginal while plays an important role in the intermediate regime.
The finite size effect induces the non-vanishing energy spacing, and prevents the inter-manifold coupling,
in the absence of the energy gaps between close-lying manifolds.
The vanishing of the energy gaps leads to the overlapping of the close-lying manifolds in the energy spectrum,
which gives rise to rich direct and avoided crossings between different manifolds.
The combined symmetries determine whether the crossing is a direct or avoided one, and the avoided crossing
can be explored for state preparations and manipulations, through interaction quench between different
avoided crossings.

Our results based on the single-hole filling can be directly generalized to different systems, 
e.g. with more hole fillings or under different external potentials. Doping more holes will leads to the degeneracy between
manifolds, and could give rise to more flexible engineering manners. The modified t-J model also suggests the possibility
of the manipulation of the Heisenberg chain in the spin sector by the holes in the charge sector, which could contribute
to the investigation of the quantum magnetization with ultracold spinor atoms.

\begin{acknowledgments}
The authors would like to acknowledge L. You for inspiring discussions. This work was supported by the National Natural Science Foundation of China (Grants No. 11625417, No. 11604107, No. 91636219, and No. 11727809).
\end{acknowledgments}

\bibliography{references}


\end{document}